\documentclass[sigconf]{acmart}

\usepackage{balance}
\usepackage{xcolor} 
\usepackage{hyperref} 
\usepackage{graphicx}
\usepackage{subcaption}
\hypersetup{
    colorlinks=true,
    linkcolor=blue, 
    urlcolor=blue,  
    citecolor=blue  
}
\AtBeginDocument{%
  }

\setcopyright{acmlicensed}
\copyrightyear{2018}
\acmYear{2018}
\acmDOI{XXXXXXX.XXXXXXX}
\acmConference[Conference acronym 'XX]{Make sure to enter the correct
  conference title from your rights confirmation emai}{June 03--05,
  2018}{Woodstock, NY}
\acmISBN{978-1-4503-XXXX-X/18/06}

\copyrightyear{2025}
\acmYear{2025}
\setcopyright{cc}
\setcctype{by}
\acmConference[WWW Companion '25]{Companion Proceedings of the ACM Web Conference 2025}{April 28-May 2, 2025}{Sydney, NSW, Australia}
\acmBooktitle{Companion Proceedings of the ACM Web Conference 2025 (WWW Companion '25), April 28-May 2, 2025, Sydney, NSW, Australia}
\acmDOI{10.1145/3701716.3715203}
\acmISBN{979-8-4007-1331-6/2025/04}

\begin{document}

\title{Use of Air Quality Sensor Network Data for Real-time Pollution-Aware POI Suggestion}


\author{Giuseppe Fasano}
\email{giuseppe.fasano@poliba.it}
\affiliation{
  \institution{Polytechnic University of Bari}
  \city{Bari}
  \country{Italy}
}

\author{Yashar Deldjoo}
\email{yashar.deldjoo@poliba.it}
\affiliation{
  \institution{Polytechnic University of Bari}
  \city{Bari}
  \country{Italy}
}

\author{Tommaso di Noia}
\email{tommaso.dinoia@poliba.it}
\affiliation{
  \institution{Polytechnic University of Bari}
  \city{Bari}
  \country{Italy}
}


\author{Bianca Lau}
\affiliation{
  \institution{A.U.G. Signals Ltd.}
  \city{Toronto}
  \country{Canada}
}

\author{Sina Adham-Khiabani}
\affiliation{
  \institution{A.U.G. Signals Ltd.}
  \city{Toronto}
  \country{Canada}
}

\author{Eric Morris}
\email{eric.morris@augsignals.com}
\affiliation{
  \institution{A.U.G. Signals Ltd.}
  \city{Toronto}
  \country{Canada}
}

\author{Xia Liu}
\affiliation{
  \institution{A.U.G. Signals Ltd.}
  \city{Toronto}
  \country{Canada}
}


\author{Ganga Chinna Rao Devarapu}
\email{chinna.devarapu@mtu.ie}
\affiliation{
  \institution{Munster Technological University}
  \city{Cork}
  \country{Ireland}
}

\author{Liam O'Faolain}
\affiliation{
  \institution{Munster Technological University}
  \city{Cork}
  \country{Ireland}
}

\renewcommand{\shortauthors}{Giuseppe Fasano et al.}

\begin{abstract}

This demo paper introduces \texttt{AirSense-R}, a privacy-preserving mobile application that delivers real-time, pollution-aware recommendations for urban points of interest (POIs). By merging live air quality data from AirSENCE sensor networks in Bari (Italy) and Cork (Ireland) with user preferences, the system enables health-conscious decision-making. It employs collaborative filtering for personalization, federated learning for privacy, and a prediction engine to detect anomalies and interpolate sparse sensor data. The proposed solution adapts dynamically to urban air quality while safeguarding user privacy. The \textbf{code} and demonstration \textbf{video} are available at \href{https://github.com/AirtownApp/Airtown-Application.git}{\textcolor{blue}{https://github.com/AirtownApp/Airtown-Application.git}}.
\end{abstract}



\maketitle

\section{Introduction}

\noindent \textbf{Background.} Urban air pollution is a major global concern. The World Health Organization (WHO) reports that 99\% of the world’s population breathes air that exceeds its recommended limits \cite{who_airpollution}. This issue is particularly acute in metropolitan areas, where vehicular and industrial emissions prevail. In this setting, point-of-interest (POI) recommendation systems can enhance urban experiences by offering health-conscious options.

\begin{figure}[!t]
    \centering
    \includegraphics[width=0.880\linewidth]{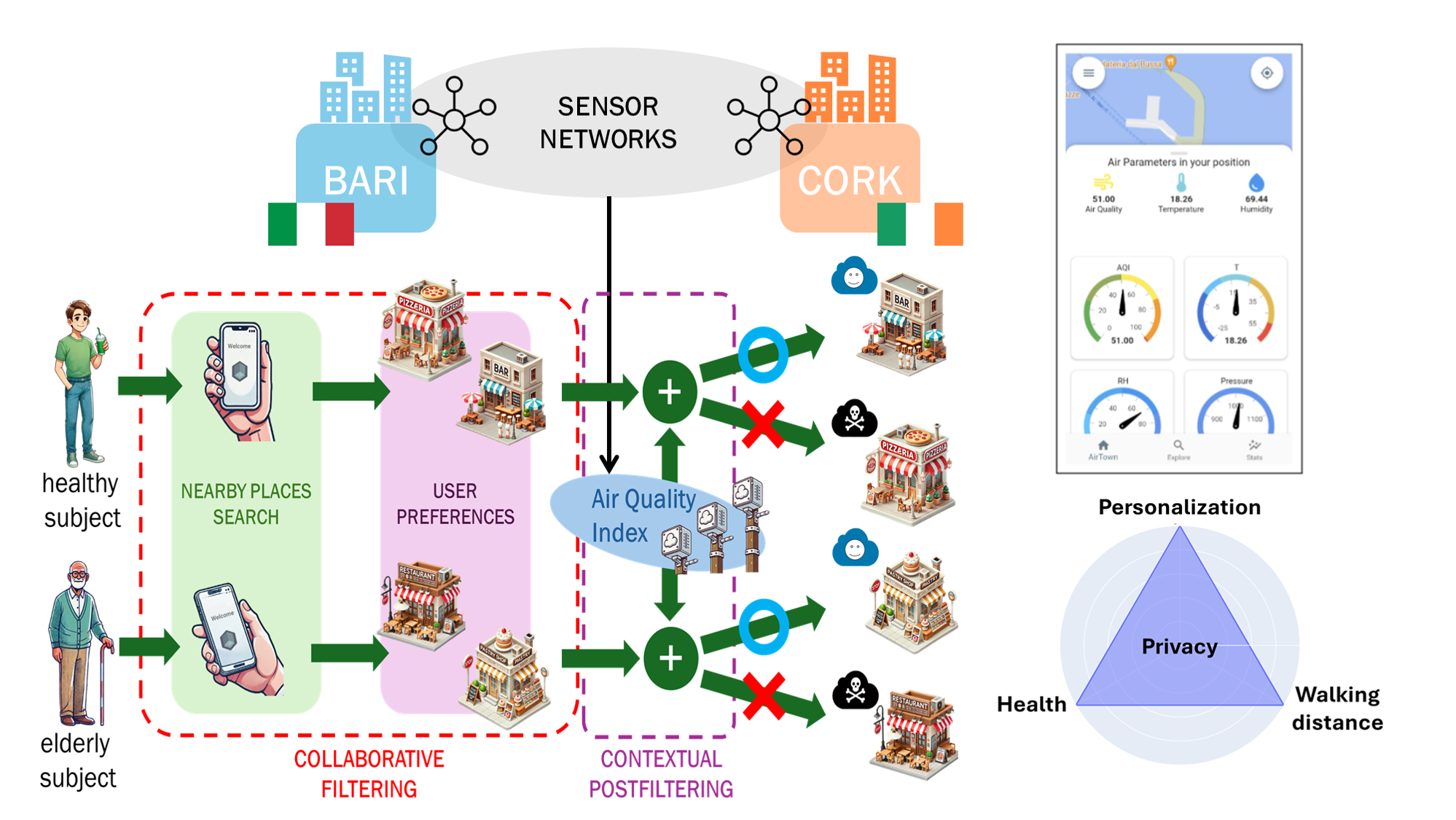}
  \caption{\texttt{AirTown} integrates real-time data from sensors installed in Bari and Cork, with user preferences to provide personalized, pollution-aware POI recommendations.}

    \label{POI_recsys_scheme}
\end{figure}

\noindent \textbf{Air Sensor Networks in Bari and Cork.} Over the past decade, several European citizen science initiatives—such as EveryAware \cite{everyaware}, Citi-Sense-MOB \cite{Castell2016,SCHNEIDER2017234}, and OpenSense \cite{MUELLER2016171}—have employed portable sensors to deliver real-time air quality data via web and mobile apps. Additionally, stationary monitoring efforts have been made, including the deployment of 9 KOALA sensors in Gold Coast during the 2018 Commonwealth Games \cite{KUHN2021105707} and Chicago’s Array of Things (AoT) Project \cite{2019AGUFM.A24G..04P}. AirSENCE\footnote{\href{https://airsence.com}{AirSENCE.com}} has installed 10 sensors in Bari and 8 in Cork since July 2022, covering areas like schools, residential neighborhoods, and city centers to capture diverse urban air quality data.

\noindent \textbf{Contribution.} This paper introduces \texttt{AirSense-R}, a system that provides real-time, personalized POI recommendations based on current air quality. By integrating predictive analytics with user preferences, it supports healthier choices in urban environments. The system comprises: (i) the \texttt{AirSense-R} Prediction Engine (see Section \ref{sec:poll_sys}) and (ii) the \texttt{AirSense-R} Recommendation Engine (\texttt{AirTown}) (see Section \ref{sec:poll_rec}).

\begin{itemize}
    \item The \texttt{AirSense-R} \textbf{Prediction} system forecasts pollutant levels using historical data from the AirSENCE networks. Methods such as FBProphet are used to capture patterns, detect anomalies, and interpolate data in areas with sparse sensor coverage.
    \item The \texttt{AirSense-R} \textbf{Recommendation} Engine, \texttt{AirTown}, is a mobile application that combines real-time pollution data with user preferences to deliver health-conscious POI recommendations. Figure \ref{POI_recsys_scheme} illustrates how the app integrates personalization, real-time data, privacy preservation, and distance considerations.
\end{itemize}

\begin{figure} [!t]
    \centering
    \includegraphics[width=1\linewidth]{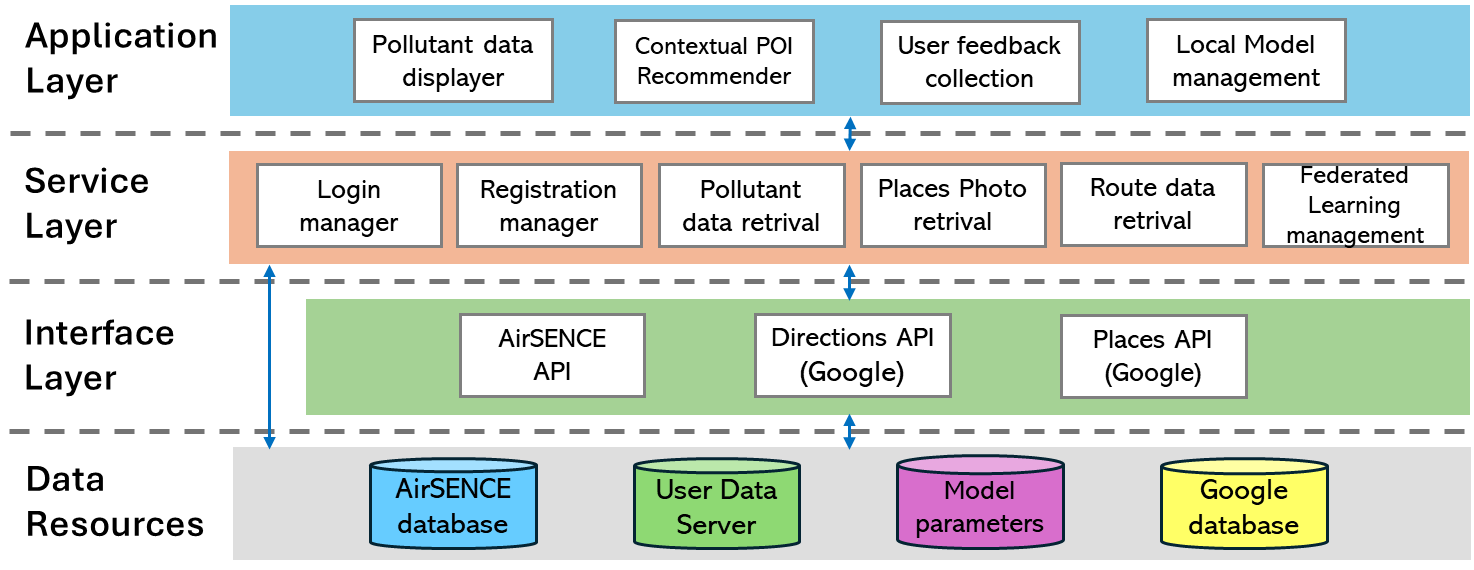}
    \caption{The architecture of the proposed system.}
    \label{FL_architecture}
\end{figure}
\section{Proposed System}
As shown in Figure \ref{FL_architecture}, the proposed system is designed to leverage a client-server model featuring four layers: Application, Service, Interface, and Data Resources Layer.


\textbf{Application Layer.} This layer serves as the user interface on mobile smartphones for \texttt{AirTown} and performs POI suggestions that integrate real-time air quality information. The suggestions are computed by the Recommendation Engine (cf. Section \ref{sec:poll_rec}). The Application Layer collects data about user preferences, but it never shares user data since the recommendation model is trained through the Federated Learning (FL) approach, allowing for user privacy.

\textbf{Service Layer.} Hosted on a server, the Service Layer manages back-end processes such as login, registration, and data retrieval for the Application Layer. It orchestrates federated learning rounds.

\textbf{Interface and Data Resources Layer.} Interface Layer comprises APIs linking the Service Layer with Data Resources. The AirSENCE API provides localized air quality data, while Directions and Places APIs enable navigation and location information. The Google Database manages Google-related data (Photos, place information, routing data), and the AirSENCE Database stores air quality measurements by sensor networks in Bari and Cork.

\subsection*{AirSENCE Sensor Network.}
A key feature of \texttt{AirTOWN} is the integration of real-time air quality data. After sensor data collection, the \textcolor{black}{Air Quality Index} (AQI) is calculated to assess air quality in real time. Effective integration relies on a robust sensor network. AirSENCE utilizes a multi-parameter sensor array with ML-based signal processing and data fusion, forming an industrial Internet of Things network. Ambient air is actively sampled, with measurements averaged every minute using on-board processing and storage before transmission to a host server. Pollutants measured include CO, NO, NO\textsubscript{2}, O\textsubscript{3}, SO\textsubscript{2}, PM\textsubscript{1}, PM\textsubscript{2.5}, PM\textsubscript{10}, temperature, humidity, and pressure. Data from two AirSENCE networks in Bari (Italy) and Cork (Ireland), over two years demonstrate the effectiveness of distributing multiple monitoring devices across diverse environments such as schools, residential areas, ferry docks, and city centers to capture urban air quality.


\begin{figure}[!t]
    \centering
    \includegraphics[width=0.75\linewidth]{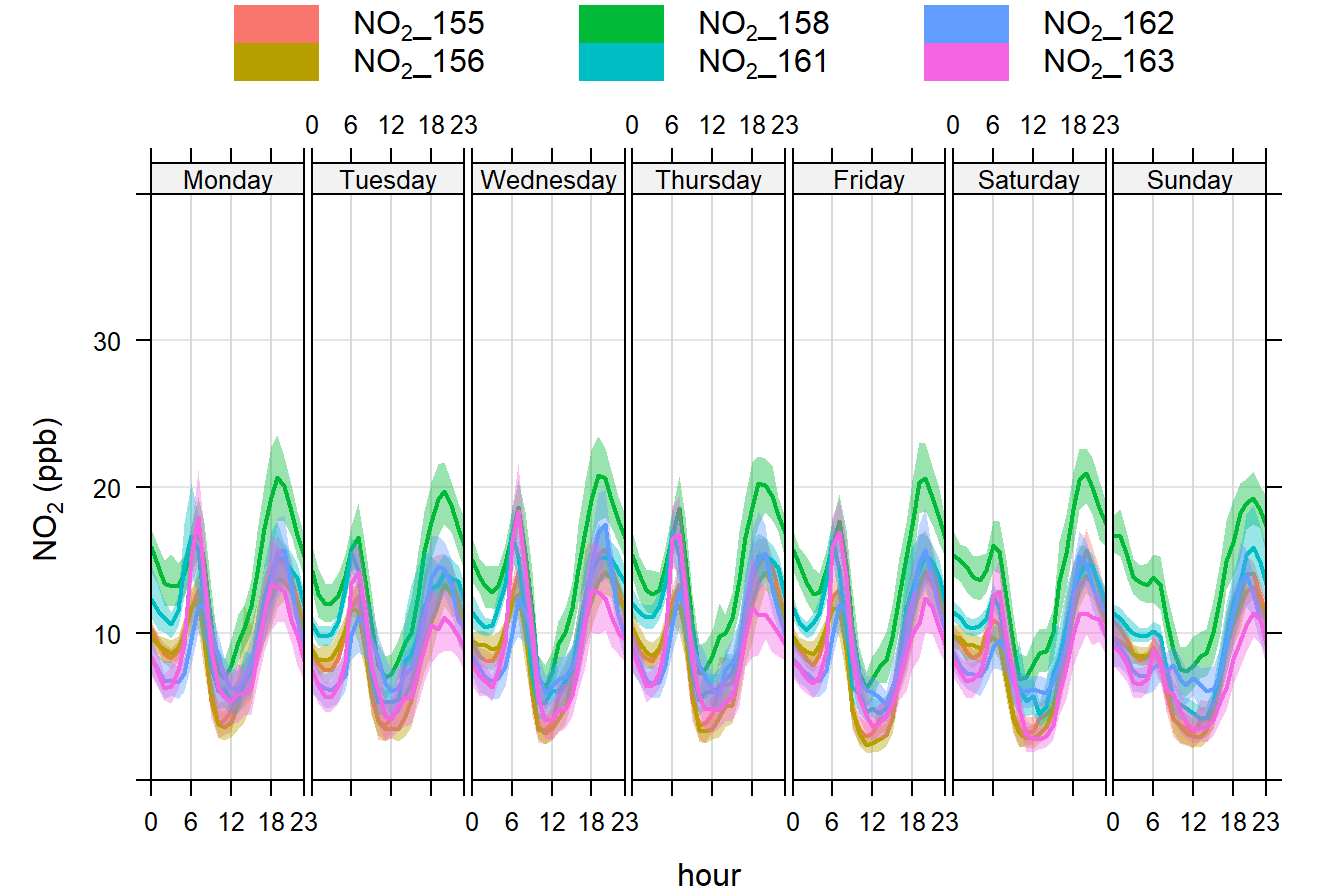}
    \caption{Weekly diurnal trends of NO\textsubscript{2} for six AirSENCE devices in Bari, Italy 2023.}
    \label{simulations2}
\end{figure}

\begin{figure}[h]
    \centering
    \includegraphics[width=0.750\linewidth]{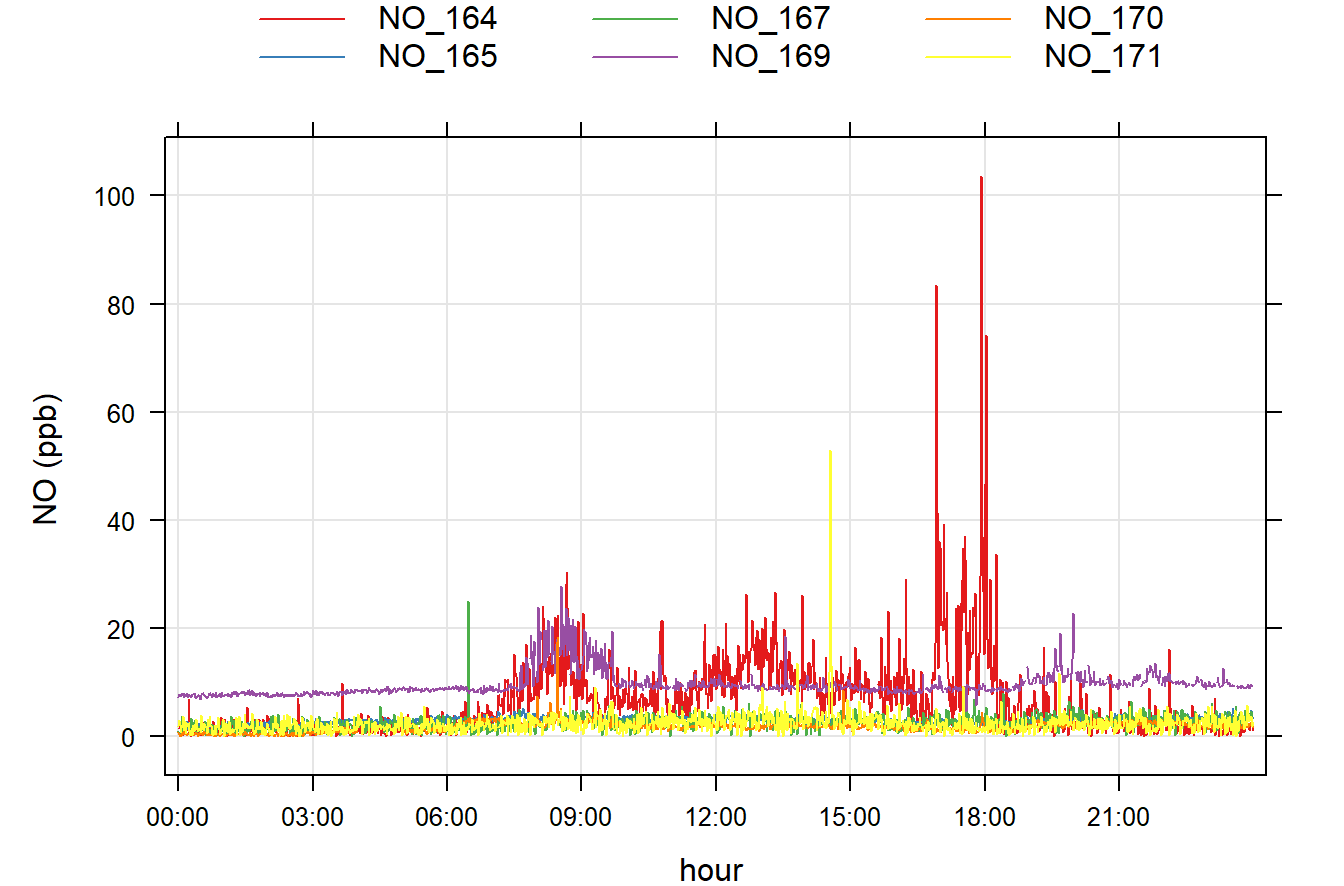}
    \caption{NO readings for six AirSENCE devices in Cork, Ireland on January 29, 2024.}
    \label{simulations3}
\end{figure}

For instance, Figure \ref{simulations2} displays the weekly diurnal patterns of NO\textsubscript{2} from six AirSENCE devices in Bari, Italy, in 2023. These devices were located near a daycare, an industrial parking lot, a residential cul-de-sac, and a ferry dock port. While all devices showed similar weekly trends, NO\textsubscript{2} concentrations varied significantly by location, primarily due to vehicle emissions. Higher concentrations occurred during rush hours around 6 a.m. and 6:30 p.m., with the highest levels at device 158 in a residential area, likely reflecting increased traffic. This example underscores the advantage of multiple monitoring devices in providing a more accurate local air quality representation compared to single monitoring stations.

Additionally, multiple real-time monitoring devices can detect sudden, localized pollution events. Figure  \ref{simulations3} illustrates a notable spike in NO levels by device 164 around 6 p.m., which was not observed by nearby devices. This highlights how real-time data can identify abrupt air quality changes that static data might miss, enhancing the ability to offer accurate and timely recommendations based on current air conditions.

\vspace{-2mm}

\subsection*{AirSENCE Prediction Engine}
\label{sec:poll_sys}

Predicting and modeling air pollutant concentrations enhance air quality analysis by identifying patterns and anomalies in real-time data. Air quality data shows long-term trends, daily/weekly cycles from human activities, and seasonal variations influenced by climate factors. Techniques such as Seasonal-Trend Decomposition using Loess (STL) and Fourier Transforms separate time series into trend, seasonality, and residuals. Advanced models like ARIMA, Long Short-Term Memory (LSTM)\cite{ref2}, and FBProphet capture complex interactions. FBProphet is particularly effective due to its flexibility with irregular data, non-linear trend modeling, and support for seasonality and holidays, as demonstrated by its accurate NO forecasts in Bari, Italy.

Pollution events—like wildfire-induced particulate spikes or elevated NO\textsubscript{2} during rush hours—are detected through residual analysis that highlights deviations from expected patterns. This anomaly detection method, which considers factors like time, season, and location, offers more precise assessments than traditional AQI methods. Furthermore, prediction models enable spatial interpolation in areas with sparse sensor coverage. Integrating the AirSENCE prediction engine with a real-time monitoring network facilitates early detection of pollution events and provides localized air quality data for pollution-aware POI suggestions.

\begin{figure}[h]
    \centering
    \includegraphics[width=0.9\linewidth]{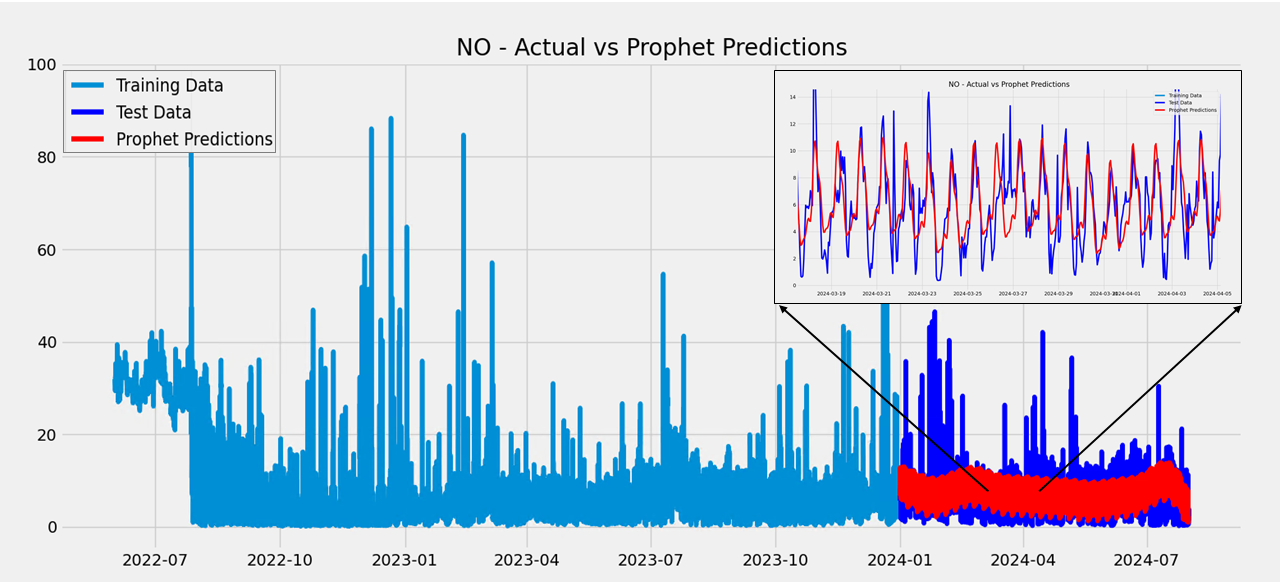}
    \caption{NO pattern prediction using FBProphet and AirSENCE device 161 in Bari, Italy.}
    \label{simulations}
\end{figure}

\subsection*{Recommendation Engine}
We developed a health-oriented POI recommendation system that balances personalization, pollution levels, distance, and privacy. Personalization helps users discover new items aligned with their tastes, while integrating pollution awareness encourages healthier choices. To promote walking exploration, only POIs within a given radius (e.g., 1\,km) are considered. Finally, user privacy is preserved by keeping preferences and mobility data local, in compliance with GDPR~\cite{GDPR}.

As illustrated in Figure~\ref{POI_recsys_scheme}, the system first obtains nearby POIs via the Places API. A Matrix Factorization (MF) model~\cite{koren2009matrix} then predicts user preferences, chosen for its transparency and suitability for resource-limited devices. Real-time AQI values are retrieved from the AirSENCE Database; missing areas are approximated with radial basis function interpolation. The final recommendation score for a POI is:
\[
S = \alpha \cdot S_{\mathrm{MF}} + (1 - \alpha) \cdot S_{\mathrm{AQI}},
\]
where $\alpha \in [0,1]$ modulates the weight of air quality in the re-ranking step.

To train the MF model, user preferences are gathered locally (e.g., through surveys) and remain on the device. We adopt a client-server Federated Learning (FL) scheme, where each client holds a private user embedding and a shared POI embedding matrix. During training, clients update both embeddings but only send the POI embedding updates to the server, which then aggregates them via Federated Averaging~\cite{mcmahan2017communication} and redistributes the aggregated updates to all clients. This process can be scheduled periodically (e.g., biweekly) to preserve privacy while continuously improving recommendations.

\section{Demonstration and Results}
\label{sec:poll_rec}

We used a dataset of 11{,}606 ratings from 8{,}982 users on 2{,}594 POIs in Bari, with each rating ranging from 1 to 5. Due to limited sensor coverage, we simulated eight virtual sensors in a $1\times1$\,km area around each user, assigning random AQI values (20--70). Two users in Aldo Moro Square were considered: 
\begin{itemize}
    \item \textbf{User 1:} A generally healthy individual. 
    \item \textbf{User 2:} An elderly person with higher sensitivity to air quality.
\end{itemize}
Each user’s preferences were locally collected to compute embeddings for personalized recommendations.

\subsubsection*{Intra-User Demonstration}
Figure~\ref{results} (red-boxed lists) shows how varying $\alpha$ shifts the recommendation focus:
\begin{itemize}
    \item $\alpha=0$ (AQI-driven): Only air quality matters.
    \item $\alpha=1$ (preference-driven): Only personal preferences matter.
    \item $\alpha=0.5$ (balanced): A mix of preferences and AQI.
\end{itemize}

\subsubsection*{Inter-User Demonstration}
Figure~\ref{results} (blue-boxed list) shows that User 2, with $\alpha=0.3$, is more cautious about AQI, receiving different recommendations than User 1. This illustrates \texttt{AirTOWN}'s flexibility in addressing individual health needs.

\subsubsection*{Federated Learning Demonstration}
\begin{figure}[H]
    \centering
    \includegraphics[width=0.95\linewidth]{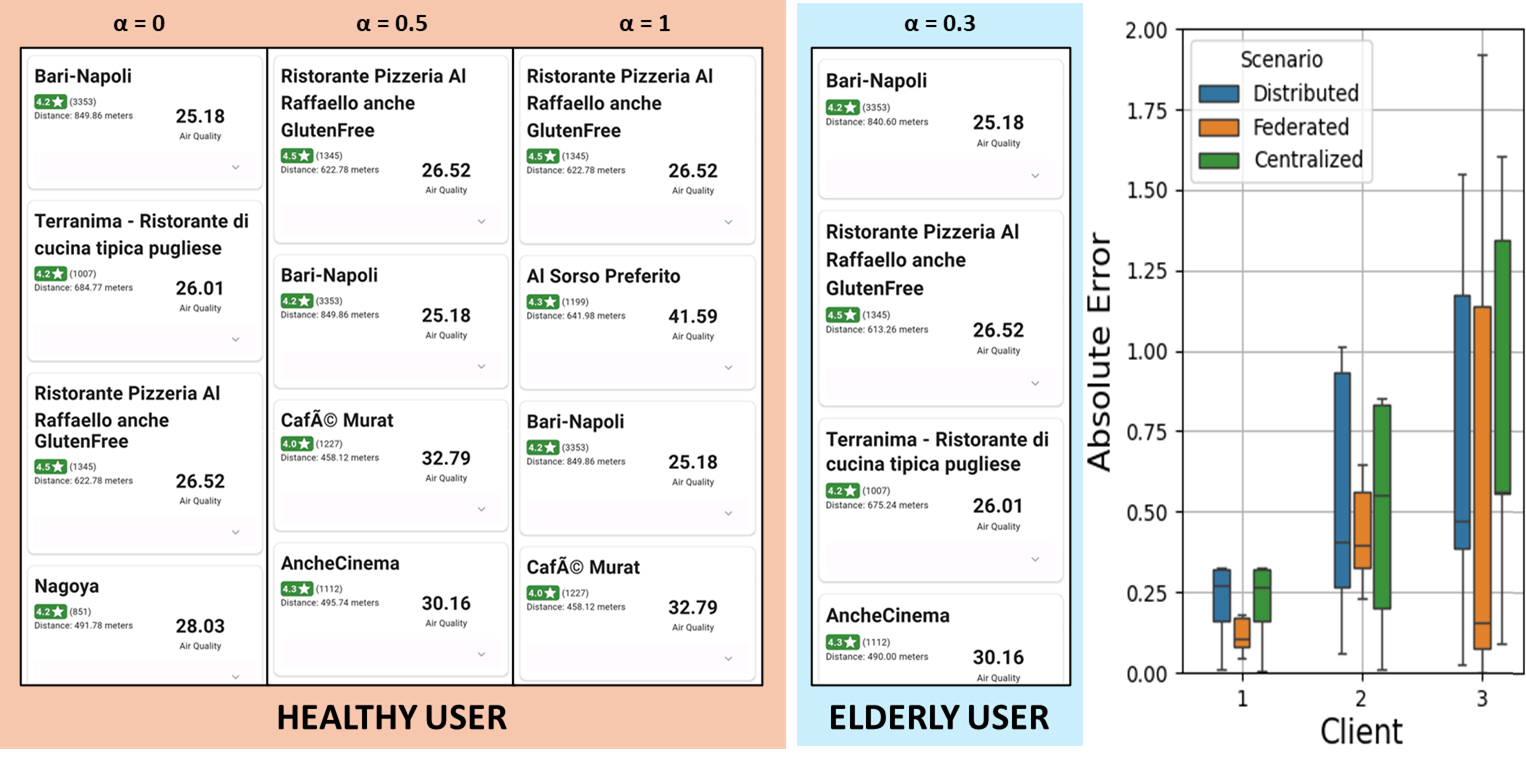}
    \caption{Demonstration results for AirTown.}
    \label{results}
\end{figure}

A baseline matrix-factorization model was first trained on the entire dataset except for three users with the highest number of ratings. Those users were then involved in a second training step under three scenarios: centralized, distributed, and federated. The boxplots in Figure~\ref{results} confirm that the federated model consistently achieves the lowest median absolute error per user, highlighting its effectiveness for personalized and privacy-preserving learning.

\section{Conclusion}



We presented \texttt{AirSense-R}, a mobile application that combines personalization, real-time pollution awareness, privacy, and proximity considerations to deliver health-conscious point-of-interest (POI) recommendations. By integrating collaborative filtering with federated learning, \texttt{AirSense-R}~provides users with tailored suggestions while protecting the privacy of personal data. Initial experiments highlight the effectiveness of the Mobile App. in balancing user preferences with real-time air quality data, making it a valuable tool for urban navigation that promotes healthier choices. 

\textcolor{black}{\texttt{AirSense-R} reserves room for improvement. Currently, the recommendation engine employs only AQI as a source of context for pollution, but data such as temperature or humidity could enhance the usage of the system.
Furthermore, the prediction engine could be used to predict the best time in the day to reach a suggested POI, improving \texttt{AirTOWNS}'s applicability. Further investigation on the performance of the application in realistic conditions will be held, alongside user studies to gauge users' perception of the application.}

This paper positions \textcolor{black}{\texttt{AirSense-R}} within the broader landscape of trustworthy recommender systems \cite{deldjoo2022survey2,deldjoo2024understanding,deldjoo2025cfairllm,nazary2025poison}, highlighting its emphasis on privacy preservation, transparency, and health-conscious decision-making \cite{10.1145/2792838.2796554,deldjoo2018content}. Additionally, we aim to incorporate insights from recent advancements in generative recommender models \cite{biancofiore2024interactive,deldjoo2024review,deldjoo2024recommendation}, which underscore the potential for dynamic, user-adaptive recommendations.

\subsubsection*{Acknowledgments}
This work was partially supported by the following projects: PASSEPARTOUT, LUTECH DIGITALE 4.0, P+ARTS, 2022LKJWHC - TRex-SE: Trustworthy Recommenders for Software Engineers, 2022ZLL7MW - Conversational Agents: Mastering, Evaluating, Optimizing.

\bibliographystyle{ACM-Reference-Format}
\balance
\bibliography{main}


\begin{thebibliography}{20}


\ifx \showCODEN    \undefined \def \showCODEN     #1{\unskip}     \fi
\ifx \showDOI      \undefined \def \showDOI       #1{#1}\fi
\ifx \showISBNx    \undefined \def \showISBNx     #1{\unskip}     \fi
\ifx \showISBNxiii \undefined \def \showISBNxiii  #1{\unskip}     \fi
\ifx \showISSN     \undefined \def \showISSN      #1{\unskip}     \fi
\ifx \showLCCN     \undefined \def \showLCCN      #1{\unskip}     \fi
\ifx \shownote     \undefined \def \shownote      #1{#1}          \fi
\ifx \showarticletitle \undefined \def \showarticletitle #1{#1}   \fi
\ifx \showURL      \undefined \def \showURL       {\relax}        \fi
\providecommand\bibfield[2]{#2}
\providecommand\bibinfo[2]{#2}
\providecommand\natexlab[1]{#1}
\providecommand\showeprint[2][]{arXiv:#2}

\bibitem[who(2023)]%
        {who_airpollution}
 \bibinfo{year}{Accessed June 16, 2023}\natexlab{}.
\newblock \bibinfo{title}{Air Pollution}.
\newblock \bibinfo{howpublished}{\url{https://www.who.int/health-topics/air-pollution}}.
\newblock


\bibitem[Biancofiore et~al\mbox{.}(2024)]%
        {biancofiore2024interactive}
\bibfield{author}{\bibinfo{person}{Giovanni~Maria Biancofiore}, \bibinfo{person}{Yashar Deldjoo}, \bibinfo{person}{Tommaso~Di Noia}, \bibinfo{person}{Eugenio Di~Sciascio}, {and} \bibinfo{person}{Fedelucio Narducci}.} \bibinfo{year}{2024}\natexlab{}.
\newblock \showarticletitle{Interactive Question Answering Systems: Literature Review}.
\newblock \bibinfo{journal}{\emph{Comput. Surveys}} \bibinfo{volume}{56}, \bibinfo{number}{9} (\bibinfo{year}{2024}), \bibinfo{pages}{1--38}.
\newblock


\bibitem[Castell et~al\mbox{.}(2016)]%
        {Castell2016}
\bibfield{author}{\bibinfo{person}{N{\'u}ria Castell}, \bibinfo{person}{Hai-Ying Liu}, \bibinfo{person}{Franck~R. Dauge}, \bibinfo{person}{Mike Kobernus}, \bibinfo{person}{Arne~J. Berre}, \bibinfo{person}{Josef Noll}, \bibinfo{person}{Erol Cagatay}, {and} \bibinfo{person}{Reidun Gangdal}.} \bibinfo{year}{2016}\natexlab{}.
\newblock \bibinfo{booktitle}{\emph{Supporting Sustainable Mobility Using Mobile Technologies and Personalized Environmental Information: The Citi-Sense-MOB Approach in Oslo, Norway}}.
\newblock \bibinfo{publisher}{Springer International Publishing}, \bibinfo{address}{Cham}, \bibinfo{pages}{199--218}.
\newblock
\showISBNx{978-3-319-23455-7}
\urldef\tempurl%
\url{https://doi.org/10.1007/978-3-319-23455-7_11}
\showDOI{\tempurl}


\bibitem[Deldjoo(2024)]%
        {deldjoo2024understanding}
\bibfield{author}{\bibinfo{person}{Yashar Deldjoo}.} \bibinfo{year}{2024}\natexlab{}.
\newblock \showarticletitle{{Understanding Biases in ChatGPT-based Recommender Systems: Provider Fairness, Temporal stability, and Recency}}.
\newblock \bibinfo{journal}{\emph{ACM Transactions on Recommender Systems}} (\bibinfo{year}{2024}).
\newblock


\bibitem[Deldjoo and Di~Noia(2025)]%
        {deldjoo2025cfairllm}
\bibfield{author}{\bibinfo{person}{Yashar Deldjoo} {and} \bibinfo{person}{Tommaso Di~Noia}.} \bibinfo{year}{2025}\natexlab{}.
\newblock \showarticletitle{{CFaiRLLM: Consumer Fairness Evaluation in Large-Language Model Recommender System}}.
\newblock \bibinfo{journal}{\emph{ACM Transactions on Intelligent Systems and Technology (TIST)}} (\bibinfo{year}{2025}).
\newblock


\bibitem[Deldjoo et~al\mbox{.}(2024a)]%
        {deldjoo2024review}
\bibfield{author}{\bibinfo{person}{Yashar Deldjoo}, \bibinfo{person}{Zhankui He}, \bibinfo{person}{Julian McAuley}, \bibinfo{person}{Anton Korikov}, \bibinfo{person}{Scott Sanner}, \bibinfo{person}{Arnau Ramisa}, \bibinfo{person}{Ren{\'e} Vidal}, \bibinfo{person}{Maheswaran Sathiamoorthy}, \bibinfo{person}{Atoosa Kasirzadeh}, {and} \bibinfo{person}{Silvia Milano}.} \bibinfo{year}{2024}\natexlab{a}.
\newblock \showarticletitle{A Review of Modern Recommender Systems using Generative Models (Gen-RecSys)}. In \bibinfo{booktitle}{\emph{Proceedings of the 30th ACM SIGKDD Conference on Knowledge Discovery and Data Mining}}. \bibinfo{pages}{6448--6458}.
\newblock


\bibitem[Deldjoo et~al\mbox{.}(2024b)]%
        {deldjoo2024recommendation}
\bibfield{author}{\bibinfo{person}{Yashar Deldjoo}, \bibinfo{person}{Zhankui He}, \bibinfo{person}{Julian McAuley}, \bibinfo{person}{Anton Korikov}, \bibinfo{person}{Scott Sanner}, \bibinfo{person}{Arnau Ramisa}, \bibinfo{person}{Rene Vidal}, \bibinfo{person}{Maheswaran Sathiamoorthy}, \bibinfo{person}{Atoosa Kasrizadeh}, \bibinfo{person}{Silvia Milano}, {et~al\mbox{.}}} \bibinfo{year}{2024}\natexlab{b}.
\newblock \showarticletitle{Recommendation with generative models}.
\newblock \bibinfo{journal}{\emph{arXiv preprint arXiv:2409.15173}} (\bibinfo{year}{2024}).
\newblock


\bibitem[Deldjoo et~al\mbox{.}(2022)]%
        {deldjoo2022survey2}
\bibfield{author}{\bibinfo{person}{Yashar Deldjoo}, \bibinfo{person}{Dietmar Jannach}, \bibinfo{person}{Alejandro Bellogin}, \bibinfo{person}{Alessandro Difonzo}, {and} \bibinfo{person}{Dario Zanzonelli}.} \bibinfo{year}{2022}\natexlab{}.
\newblock \showarticletitle{A survey of research on fair recommender systems}.
\newblock \bibinfo{journal}{\emph{CoRR}} (\bibinfo{year}{2022}).
\newblock


\bibitem[Deldjoo et~al\mbox{.}(2018)]%
        {deldjoo2018content}
\bibfield{author}{\bibinfo{person}{Yashar Deldjoo}, \bibinfo{person}{Markus Schedl}, \bibinfo{person}{Paolo Cremonesi}, \bibinfo{person}{Gabirella Pasi}, {et~al\mbox{.}}} \bibinfo{year}{2018}\natexlab{}.
\newblock \showarticletitle{Content-based multimedia recommendation systems: definition and application domains}. In \bibinfo{booktitle}{\emph{Italian Information Retrieval Workshop}}. \bibinfo{pages}{1--4}.
\newblock


\bibitem[{European Commission}(2016)]%
        {GDPR}
\bibfield{author}{\bibinfo{person}{{European Commission}}.} \bibinfo{year}{2016}\natexlab{}.
\newblock \bibinfo{title}{Regulation ({EU}) 2016/679 of the {European} {Parliament} and of the {Council} of 27 {April} 2016 on the protection of natural persons with regard to the processing of personal data and on the free movement of such data, and repealing {Directive} 95/46/{EC} ({General} {Data} {Protection} {Regulation}) ({Text} with {EEA} relevance)}.
\newblock
\newblock
\urldef\tempurl%
\url{https://eur-lex.europa.eu/eli/reg/2016/679/oj}
\showURL{%
\tempurl}


\bibitem[Ge et~al\mbox{.}(2015)]%
        {10.1145/2792838.2796554}
\bibfield{author}{\bibinfo{person}{Mouzhi Ge}, \bibinfo{person}{Francesco Ricci}, {and} \bibinfo{person}{David Massimo}.} \bibinfo{year}{2015}\natexlab{}.
\newblock \showarticletitle{Health-aware Food Recommender System}. In \bibinfo{booktitle}{\emph{Proceedings of the 9th ACM Conference on Recommender Systems}} (Vienna, Austria) \emph{(\bibinfo{series}{RecSys '15})}. \bibinfo{publisher}{Association for Computing Machinery}, \bibinfo{address}{New York, NY, USA}, \bibinfo{pages}{333–334}.
\newblock
\showISBNx{9781450336925}
\urldef\tempurl%
\url{https://doi.org/10.1145/2792838.2796554}
\showDOI{\tempurl}


\bibitem[Koren et~al\mbox{.}(2009)]%
        {koren2009matrix}
\bibfield{author}{\bibinfo{person}{Yehuda Koren}, \bibinfo{person}{Robert Bell}, {and} \bibinfo{person}{Chris Volinsky}.} \bibinfo{year}{2009}\natexlab{}.
\newblock \showarticletitle{Matrix factorization techniques for recommender systems}.
\newblock \bibinfo{journal}{\emph{Computer}} \bibinfo{volume}{42}, \bibinfo{number}{8} (\bibinfo{year}{2009}), \bibinfo{pages}{30--37}.
\newblock


\bibitem[Kuhn et~al\mbox{.}(2021)]%
        {KUHN2021105707}
\bibfield{author}{\bibinfo{person}{Tara Kuhn}, \bibinfo{person}{Rohan Jayaratne}, \bibinfo{person}{Phong~K. Thai}, \bibinfo{person}{Bryce Christensen}, \bibinfo{person}{Xiaoting Liu}, \bibinfo{person}{Matthew Dunbabin}, \bibinfo{person}{Riki Lamont}, \bibinfo{person}{Isak Zing}, \bibinfo{person}{David Wainwright}, \bibinfo{person}{Christian Witte}, \bibinfo{person}{Donald Neale}, {and} \bibinfo{person}{Lidia Morawska}.} \bibinfo{year}{2021}\natexlab{}.
\newblock \showarticletitle{Air quality during and after the Commonwealth Games 2018 in Australia: Multiple benefits of monitoring}.
\newblock \bibinfo{journal}{\emph{Journal of Aerosol Science}}  \bibinfo{volume}{152} (\bibinfo{year}{2021}), \bibinfo{pages}{105707}.
\newblock
\showISSN{0021-8502}
\urldef\tempurl%
\url{https://doi.org/10.1016/j.jaerosci.2020.105707}
\showDOI{\tempurl}


\bibitem[Lorenzo(2024)]%
        {everyaware}
\bibfield{author}{\bibinfo{person}{Vittorio Lorenzo}.} \bibinfo{year}{Accessed December 4, 2024}\natexlab{}.
\newblock \bibinfo{title}{EveryAware: Enhance Environmental Awareness through Social Information Technologies. Project Final Report of EU FP7 ICT Project No. 265432}.
\newblock \bibinfo{howpublished}{\url{https://http://www.everyaware.eu/resources/deliverables/D7.3.pdf}}.
\newblock


\bibitem[McMahan et~al\mbox{.}(2017)]%
        {mcmahan2017communication}
\bibfield{author}{\bibinfo{person}{Brendan McMahan}, \bibinfo{person}{Eider Moore}, \bibinfo{person}{Daniel Ramage}, \bibinfo{person}{Seth Hampson}, {and} \bibinfo{person}{Blaise~Aguera y Arcas}.} \bibinfo{year}{2017}\natexlab{}.
\newblock \showarticletitle{Communication-efficient learning of deep networks from decentralized data}. In \bibinfo{booktitle}{\emph{Artificial intelligence and statistics}}. PMLR, \bibinfo{pages}{1273--1282}.
\newblock


\bibitem[Mueller et~al\mbox{.}(2016)]%
        {MUELLER2016171}
\bibfield{author}{\bibinfo{person}{M.D. Mueller}, \bibinfo{person}{David Hasenfratz}, \bibinfo{person}{Olga Saukh}, \bibinfo{person}{Martin Fierz}, {and} \bibinfo{person}{Christoph Hueglin}.} \bibinfo{year}{2016}\natexlab{}.
\newblock \showarticletitle{Statistical modelling of particle number concentration in Zurich at high spatio-temporal resolution utilizing data from a mobile sensor network}.
\newblock \bibinfo{journal}{\emph{Atmospheric Environment}}  \bibinfo{volume}{126} (\bibinfo{year}{2016}), \bibinfo{pages}{171--181}.
\newblock
\showISSN{1352-2310}
\urldef\tempurl%
\url{https://doi.org/10.1016/j.atmosenv.2015.11.033}
\showDOI{\tempurl}


\bibitem[Nazary et~al\mbox{.}(2025)]%
        {nazary2025poison}
\bibfield{author}{\bibinfo{person}{Fatemeh Nazary}, \bibinfo{person}{Yashar Deldjoo}, {and} \bibinfo{person}{Tommaso di Noia}.} \bibinfo{year}{2025}\natexlab{}.
\newblock \bibinfo{title}{Poison-RAG: Adversarial Data Poisoning Attacks on Retrieval-Augmented Generation in Recommender Systems}.
\newblock
\newblock
\showeprint[arxiv]{2501.11759}~[cs.IR]
\urldef\tempurl%
\url{https://arxiv.org/abs/2501.11759}
\showURL{%
\tempurl}


\bibitem[{Potosnak} et~al\mbox{.}(2019)]%
        {2019AGUFM.A24G..04P}
\bibfield{author}{\bibinfo{person}{M.~J. {Potosnak}}, \bibinfo{person}{P. {Banerjee}}, \bibinfo{person}{M.~B. {Berkelhammer}}, \bibinfo{person}{R. {Sankaran}}, \bibinfo{person}{V.~R. {Kotamarthi}}, \bibinfo{person}{R.~L. {Jacob}}, \bibinfo{person}{P.~H. {Beckman}}, \bibinfo{person}{S. {Shahkarami}}, \bibinfo{person}{D.~E. {Horton}}, \bibinfo{person}{A. {Montgomery}}, {and} \bibinfo{person}{C.~E. {Catlett}}.} \bibinfo{year}{2019}\natexlab{}.
\newblock \showarticletitle{{Array of Things: A high-density, urban deployment of low-cost air quality sensors}}. In \bibinfo{booktitle}{\emph{AGU Fall Meeting Abstracts}}, Vol.~\bibinfo{volume}{2019}. Article \bibinfo{articleno}{A24G-04}, \bibinfo{numpages}{A24G-04}~pages.
\newblock


\bibitem[Schneider et~al\mbox{.}(2017)]%
        {SCHNEIDER2017234}
\bibfield{author}{\bibinfo{person}{Philipp Schneider}, \bibinfo{person}{Nuria Castell}, \bibinfo{person}{Matthias Vogt}, \bibinfo{person}{Franck~R. Dauge}, \bibinfo{person}{William~A. Lahoz}, {and} \bibinfo{person}{Alena Bartonova}.} \bibinfo{year}{2017}\natexlab{}.
\newblock \showarticletitle{Mapping urban air quality in near real-time using observations from low-cost sensors and model information}.
\newblock \bibinfo{journal}{\emph{Environment International}}  \bibinfo{volume}{106} (\bibinfo{year}{2017}), \bibinfo{pages}{234--247}.
\newblock
\showISSN{0160-4120}
\urldef\tempurl%
\url{https://doi.org/10.1016/j.envint.2017.05.005}
\showDOI{\tempurl}


\bibitem[Waseem et~al\mbox{.}(2022)]%
        {ref2}
\bibfield{author}{\bibinfo{person}{Khawaja~Hassan Waseem}, \bibinfo{person}{Hammad Mushtaq}, \bibinfo{person}{Fazeel Abid}, \bibinfo{person}{Adnan~M. Abu-Mahfouz}, \bibinfo{person}{Asadullah Shaikh}, \bibinfo{person}{Mehmet Turan}, {and} \bibinfo{person}{Jawad Rasheed}.} \bibinfo{year}{2022}\natexlab{}.
\newblock \showarticletitle{Forecasting of Air Quality Using an Optimized Recurrent Neural Network}.
\newblock \bibinfo{journal}{\emph{Processes}} \bibinfo{volume}{10}, \bibinfo{number}{10} (\bibinfo{year}{2022}).
\newblock
\showISSN{2227-9717}
\urldef\tempurl%
\url{https://doi.org/10.3390/pr10102117}
\showDOI{\tempurl}


\end{thebibliography}

\end{document}